\documentclass[reprint,
 amsmath,amssymb,
 aps,
]{revtex4-2}
\usepackage{xcolor}
\usepackage{graphicx}
\usepackage{dcolumn}
\usepackage{bm}

\newcommand{\DK}[1]{\textcolor{blue}{#1}}
\newcommand{\DKK}[1]{\textcolor{purple}{#1}}
\newcommand{\SSS}[1]{\textcolor{red}{#1}}

\begin{document}


\title{Statistical mechanics of an active wheel rolling in circles}
\thanks{A footnote to the article title}%

\author{Shubham Sharma}
\author{Deepak Kumar}%
 \email{krdeepak@physics.iitd.ac.in}
\affiliation{%
 Department of Physics, Indian Institute of Technology Delhi, New Delhi - 110016\\
}%





\begin{abstract}
Vibrated granular matter constitutes a useful system for studying the physics of active matter. Usually, self-propulsion is induced in grains through suitable asymmetry in the particle design. In this paper, we show that a symmetrical mini wheel placed on a vibrating plate self-propels along circular trajectories, showing chiral active dynamics. The chiral activity emerges through a sequence of spontaneous symmetry breaking in the particle’s kinetics. The fact that isotropy, fore-aft, and chiral symmetries are broken spontaneously leads to distinct statistics, which include a temporal evolution involving stochastic resetting, a non-Gaussian velocity distribution with multiple peaks, broad power-law curvature distribution, and a bounded chirality probability, along with a phase transition from passive achiral to active chiral state as a function of vibration amplitude. Our study establishes the vibrated wheel as a three-state chiral active system that can serve as a model experimental system to study the non-equilibrium statistical mechanics and stochastic thermodynamics of chiral active systems and can inspire novel locomotion strategies in robotics.

\end{abstract}

\maketitle

\section{Introduction}
The complexity and rich functional response of the biological world continue to amaze and inspire us. It is now widely accepted that an important ingredient of this complexity lies in the collective behavior of active constituents capable of absorbing energy locally and converting it into systematic motion\cite{ramaswamy2010mechanics}. These active entities range in size from the microscopic to the macroscopic. Understanding their inherently out-of-equilibrium response that involves effects such as spontaneous flow, symmetry breaking and topological defect dynamics constitutes a challenging physics problem\cite{marchetti2013hydrodynamics,bowick2022symmetry,fodor2016far}, and there is intense ongoing research to develop a consistent physics framework that can help us interpret and predict their behavior\cite{gompper20202020}.   

These efforts have been greatly helped by the introduction of synthetic active systems that mimic their biological counterparts and, at the same time, allow us to perform well-controlled and systematic experiments\cite{RevModPhys.88.045006,gompper20202020}. Vibrated granular matter constitutes an important class of such synthetic active systems\cite{RevModPhys.88.045006,kumar2014flocking,arora2022motile,walsh2017noise, ramaswamy2010mechanics}. In the biological world, many active particles swim along circular trajectories showing chiral dynamics\cite{RevModPhys.88.045006,jennings1901significance,lauga2006swimming,friedrich2008stochastic}. Some synthetic systems that mimic such circle swimming behavior have also been designed and studied, however, mostly at colloidal scale \cite{PhysRevLett.110.198302,shelke2019transition} with very few granular matter examples\cite{nakata1997self,dauchot2019dynamics}. 

In this paper, we introduce a simple granular realization of a particle that shows chiral active motion 
propelling itself along circular trajectories. Our system consists of a vertically vibrated toroid, about half a centimeter in diameter and a millimeter in thickness—in other words, a simple mini wheel. It is quite remarkable that a particle with an axially symmetric geometry shows chiral active dynamics. It may be noted that in most other active systems, a symmetry breaking in the dynamics is achieved by explicitly introducing asymmetries in the particle shape or in its interaction with the environment. 
We show that underlying the observed chiral activity of our system lies a sequence of spontaneous symmetry breaking, which leads to a statistical mechanics completely different from other circle-swimming active particles studied before.

\begin{figure}[h]
\includegraphics[width=0.45\textwidth]{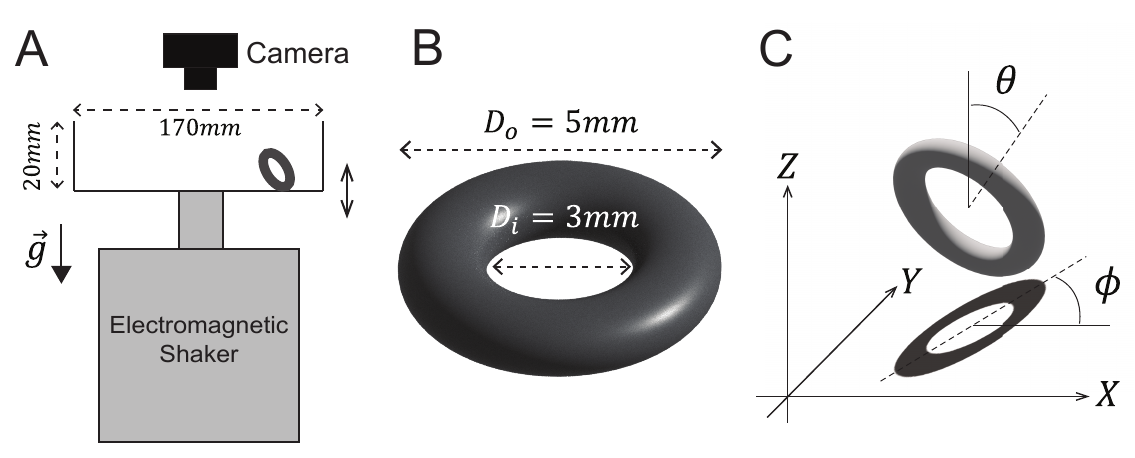}
\caption{\label{fig:setup} Experimental details. (A) Schematic of the setup. (B) Toroidal geometry of the particle. (C) Various particle coordinates used to describe the dynamics.}
\label{fig:setup}
\end{figure}

\section{Experiment}
A single torus of nitrile rubber, having outer diameter $D_o=5$ mm and inner diameter $D_i=3$ mm, is placed on a horizontal circular plate which is vibrated using an electromagnetic shaker (Pacific Dynamics, PDS-10) (Fig. \ref{fig:setup}) at an acceleration amplitude $\Gamma$ which is varied from $1.40g$ to $2.65g$, $g$ being the acceleration due to gravity. Collisions with the vibrating plate induce motion in various coordinates of the particle: $x,y,z,\theta,$ and $\phi$ (Fig. \ref{fig:setup}C). Top view of the system, involving elliptical projection of the particle, is imaged for $ 900$ s using a DSLR camera (Olympus OM-D E-M10 Mark II). 
From the images, we determine the centroid $\vec{r}=(x,y)$ of the ellipse and the angle $\phi$ between its major axis and the $x$-axis. The angle $\theta$ between the particle axis and the vertical is measured using the ratio $b/a$ of the minor axis $b$ to the major axis $a$ as described in the SI.

\section{Results and Discussion}

\subsection{Trajectories}
In Fig. \ref{fig:motion}A, B, and C, we show typical particle trajectories at $\Gamma=1.48g, 1.65g$  and $2.65g$, respectively (see also movies: S1 to S3). For small $\Gamma$ (Fig. \ref{fig:motion}A), we see localized and intermittent random motion. 
As $\Gamma$ is increased beyond a threshold value ($\Gamma_c=1.49g$), the particle's dynamics changes completely--it begins to roll on the substrate in persistent ordered motion along circular orbits spread over a distance many times the particle size, as shown in Fig. \ref{fig:motion}B. The persistent circular motion is interspersed with spells of random noisy motion. During an uninterrupted stretch of circular motion, lasting over a few seconds, the particle maintains the same sense of chirality (clockwise or anticlockwise). As we further increase $\Gamma$ (Fig. \ref{fig:motion}C), the radius of the circular orbit increases, and the trajectory approaches a straight line. The particle now collides with the walls of the container often, which disrupts the persistent circular motion. 

\subsection{Mean Squared Displacement}
\begin{figure}
    \includegraphics[width=0.48\textwidth]{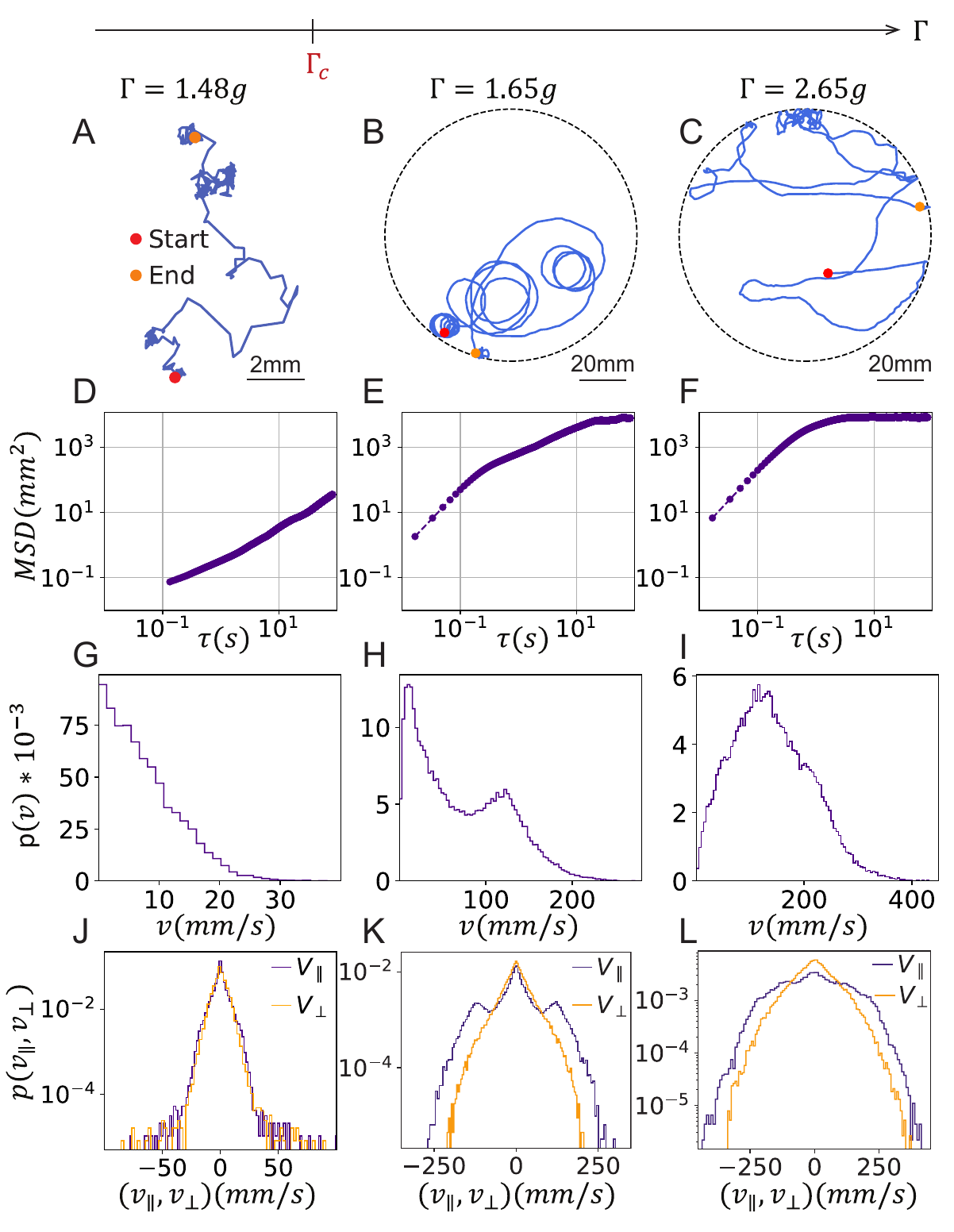}
    \caption{ Observed particle dynamics. (A), (B), and (C) Typical particle trajectory in the horizontal plane for $10$ s, (D), (E), and (F) mean squared displacement ($MSD$), (G), (H), and (I) speed distribution (J), (K), and (L) distributions of $v_\perp$ (orange) and $v_\parallel$ (purple),  for $\Gamma=1.48g$, $1.65g$, and $2.65g$, respectively. The dotted circles in (B) and (C) represent the system's boundary. }

    \label{fig:motion}
\end{figure}

We plot the mean squared displacement $MSD(\tau)$  
on log-log scale in Fig. \ref{fig:motion}D-F. For $\Gamma<\Gamma_c$ (Fig. \ref{fig:motion}D), the $MSD$ has a slope $\sim1$, characteristic of diffusive motion: $MSD(\tau)=D\tau$ with $D$ representing the diffusivity. As $\Gamma$ crosses $\Gamma_c$ (Fig. \ref{fig:motion}E), the $MSD$ increases by at least two orders of magnitude and, quite remarkably, now at small time scales, it has a slope $\sim2$, characteristic of ballistic motion, and transitions to a slope close to $1$ only at longer time scales $\tau>\tau_{MSD}$. The emergence of a ballistic regime that lasts over a macroscopic time scale is one of the hallmarks of active motion and corresponds, in the present case, to the emergence of persistent circular motion. The crossover from ballistic to diffusive behavior at a time scale $\sim \tau_{MSD}$ happens as the persistent motion is randomized due to fluctuations. At even larger values of $\tau$, we notice a saturation with $MSD\rightarrow L^2$, a finite-size effect of the container. As we increase $\Gamma$ further (Fig. \ref{fig:motion}F), the $MSD$ transitions directly from a slope $\sim 2$ to saturation, reflecting the large activity attained by the particle. 

In Fig. \ref{fig:parameters}A, we show the variation of $D$ and $\tau_{MSD}$ with $\Gamma$ (see SI for details). Both increase rapidly for $\Gamma>\Gamma_c$ but for large $\Gamma$ are affected by the finite-size effect of the container. For $\Gamma>\Gamma_c$, we obtain a measure of the particle's speed by fitting $MSD = v_{MSD}^2\tau^2$ to the initial ballistic part of the $MSD$, taking only the data corresponding to $\tau<0.1$ s (Fig. \ref{fig:parameters}B).

\subsection{Speed distribution}
\begin{figure}
    \includegraphics[width=0.5\textwidth]
    {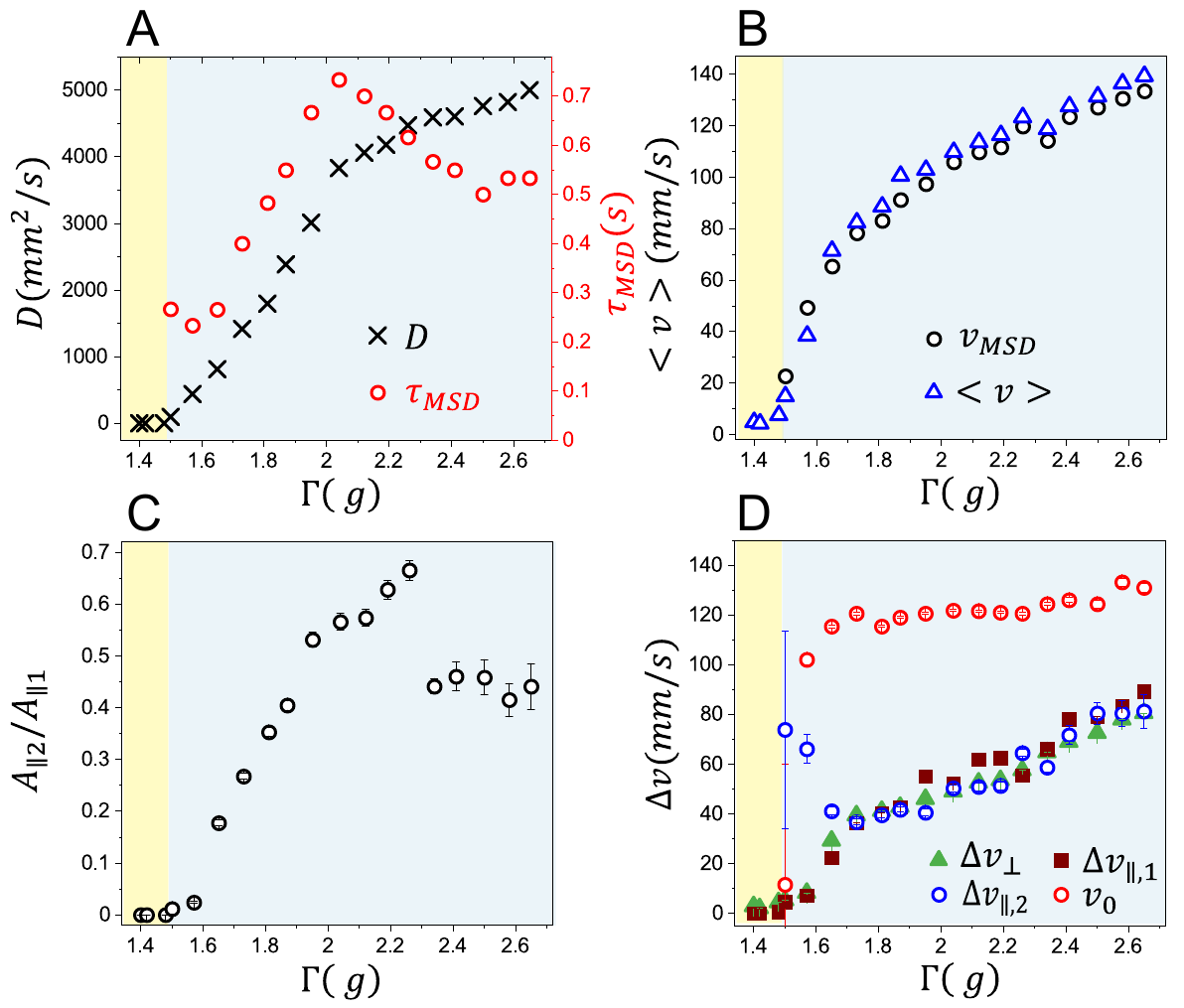}
    \caption{Effect of vibration amplitude. (A) Diffusion coefficient $D$ (black cross) and cross-over time $\tau_{MSD}$ (red open circle), (B) ballistic speed $v_{MSD}$ (black open circles) obtained from $MSD$ and mean speed $\langle v\rangle$ (blue open triangle), (C) the relative heights of the two peaks in $p(v_\parallel)$, (D) the parameters $\Delta v_{\perp}$ (filled triangle), $\Delta v_{\parallel1}$ (filled square), and $\Delta v_{\parallel2}$ (open circle) characterizing the widths of the various peaks in $p(v_\perp)$ and  $p(v_\parallel)$ and the position of the second peak $v_0$ in $p(v_\parallel)$, as a function of $\Gamma$. In all the plots, the regions $\Gamma<\Gamma_c$ and $\Gamma>\Gamma_c$ have been shaded with yellow and light blue colors, respectively. }

    \label{fig:parameters}
\end{figure}

In Fig. \ref{fig:motion}G-I, we plot the probability distribution of speed $p(v)$. 
For $\Gamma<\Gamma_c$ (Fig. \ref{fig:motion}G), $p(v)$ has a peak close to $0$ mm/s and decays quickly for larger $v$. 
As $\Gamma$ is increased beyond $\Gamma_c$ (Fig. \ref{fig:motion}H), $p(v)$ extends to a much larger speed range. But most remarkably, a new peak emerges at $v\sim120$ 
 mm/s. As we increase $\Gamma$ further (Fig. \ref{fig:motion}I), the second peak becomes more pronounced and has a longer tail, even though the peak position doesn't change significantly. 
The first peak becomes relatively insignificant and merges with the second peak. The appearance of a second peak in the speed distribution with a long tail is another signature of the emergence of activity in the system. 
The mean speed $\langle v\rangle$ varies in a manner similar to $v_{MSD}$ (Fig. \ref{fig:parameters}B).

The transition in the particle's dynamics at $\Gamma=\Gamma_c$ corresponds physically to the emergence of rolling motion, as evident in the SI movies S1 to S3, which in turn is associated with a large non-zero value of $\theta$. Hence, to bring out the correlation between rolling motion and the enhanced particle speed, we plot in Fig. \ref{fig:resetting} A-C color maps showing the joint probability distribution $p(v,\cos{\theta})$ for the three different values of $\Gamma$. These graphs show that large particle speed is correlated with the near vertical orientation of the particle, a reflection of its rolling dynamics. 

\subsection{Velocity distribution}
Active particles usually have an asymmetry in their design with a well-defined body axis along which they self-propel. In contrast, our particle has an axially symmetric shape, and for $\Gamma<\Gamma_c$, the motion is isotropic in the horizontal plane. However, for $\Gamma>\Gamma_c$, the isotropy is spontaneously broken as $\theta$ attains large values, and the particle's projection on the horizontal plane becomes elliptical with its long axis making an angle $\phi$ with $\hat{x}$. 
In order to characterize any corresponding anisotropy in the dynamics, we resolve the particle's horizontal velocity in directions parallel and perpendicular to the observed long axis: $v_{\parallel}=v_x cos(\phi)+v_ysin(\phi)$ and $v_{\perp}=-v_xsin(\phi)+v_y cos(\phi)$. Figure \ref{fig:motion}J-L show the distributions $p(v_{\perp})$ and $p(v_{\parallel})$. 
While for $\Gamma<\Gamma_c$, both $p(v_\perp)$ and $p(v_\parallel)$ are identically distributed, confirming isotropic dynamics, for $\Gamma>\Gamma_c$ they show distinctly different behaviors--$p(v_{\parallel})$ extends over a much larger velocity range and shows two more peaks symmetrically situated about the origin.

We, therefore, conclude that for $\Gamma>\Gamma_c$, the isotropy in the dynamics is spontaneously broken, and the particle begins propelling preferentially along the long axis that emerges in its projection on the horizontal plane. Further, although the elliptical projection is apolar by itself, the fore-aft symmetry is broken kinetically through the observed rolling motion. The particle continues to be in a state of such persistent motion for a few to tens of seconds, after which the motion is interrupted by a random noisy spell, followed by another spell of persistent motion. In each such spontaneous symmetry breaking event, the propulsion direction is randomly selected. Therefore, over a large time scale, $v_\parallel$ shows a symmetrical distribution about the origin.   

It may be noted that the velocity distributions, when plotted on a log-linear scale, as in Fig. \ref{fig:motion}J-L, appear linear, implying, quite remarkably, an exponential distribution. The following functions fit the data well (see SI): 

\begin{equation}
P(v_{\perp})=A_{\perp}e^{-\lvert v_{\perp}\rvert/{\Delta v_{\perp}}}
\label{eq:vperp}
\end{equation}

\begin{equation}
P(v_{\parallel})=A_{\parallel1}e^{-\lvert v_{\parallel}\rvert/{\Delta v_{\parallel1}}}+A_{\parallel2}e^{-\lvert v_{\parallel}-v_0\rvert/{\Delta v_{\parallel2}}}
\label{eq:vpar}
\end{equation}
Using the values of the fitting parameters, we plot $A_{\parallel 2}/A_{\parallel 1}$ in Fig. \ref{fig:parameters}C, and $\Delta v_\perp$, $\Delta v_{\parallel 1}$, $\Delta v_{\parallel 2}$ and $v_0$ in Fig. \ref{fig:parameters}D, respectively. 

\subsection{Stochastic resetting}
It is quite remarkable that the velocity components show Laplacian distribution. Random motion of particles, e.g., Brownian motion, usually shows Gaussian velocity distribution as a consequence of the central limit theorem\cite{kardar2007statistical}. Active Brownian particles, including many active granular systems \cite{walsh2017noise,Workamp2018} also show a Gaussian velocity distribution, albeit with a mean value shifted from zero in the self-propulsion direction. 

\begin{figure}
    \includegraphics[width=0.5\textwidth]
  {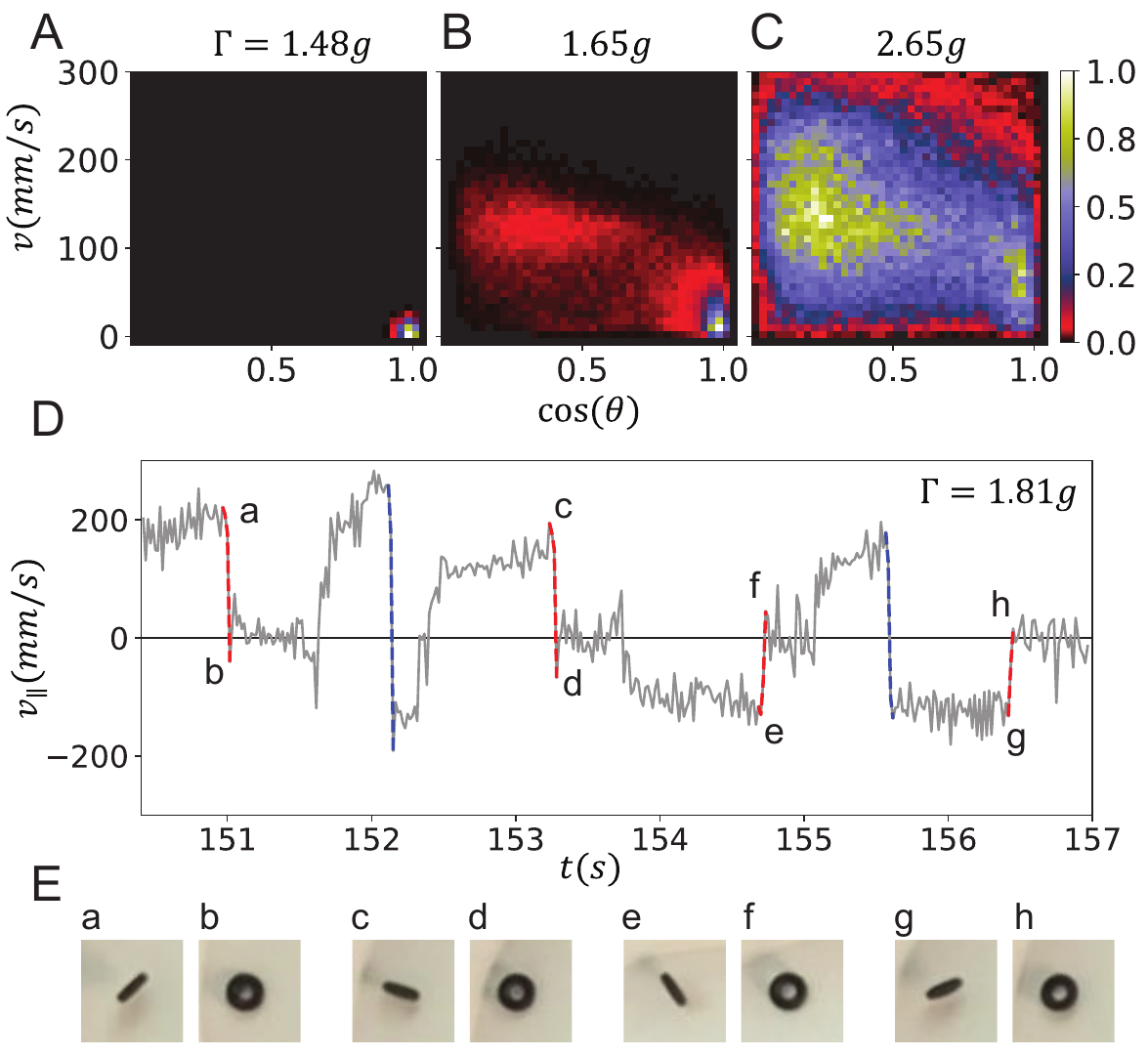}
    \caption{Stochastic resetting. (A), (B), and (C) Heat maps of the joint probability distribution $p(\cos{\theta}, v)$ for $\Gamma=1.48g$, $1.65g$, and $2.65g$ respectively. The emergence of larger speed for $\Gamma>\Gamma_c$ corresponds to rolling motion as reflected in the larger values of $\theta$.  (B) Typical time evolution of $v_\parallel$ showing stochastic resetting. The red and blue dashed lines highlight abrupt events that bring $v_\parallel$ down to zero or change the sign of $v_\parallel$ while taking its value close to $\pm v_0$, respectively. (E) Images showing the orientations of the particle at instants marked as a, b, c, d, e, f, g, and h in (D). }

    \label{fig:resetting}
\end{figure}

A plot of $ v_{\parallel}(t)$ (Fig. \ref{fig:resetting}D) helps us understand the origin of this unusual behavior. We notice that $v_{\parallel}$ evolves slowly from zero towards larger (positive or negative) values with some  occasional sudden jumps, that bring $v_{\parallel}$ down to zero again, due to collision with the walls or a random fluctuation. The particle images show that before such an event, the particle is rolling persistently, and it falls down to lie flat after it(Fig. \ref{fig:resetting}E a-h and movie S4). Sometimes, we also see that such an event leads to flipping the sign of $v_\parallel$ and to taking its magnitude close to $v_0$. 

Such a punctuated time-evolution where the variable is brought back to a particular value after randomly varying periods of time is reminiscent of stochastic resetting, a subject of intense recent research\cite{evans2020stochastic,Roichman2020,Kundu_2024}. One of the hallmarks of the stochastic resetting process is that the random variable involved shows a Laplacian distribution with a symmetric cusp at the resetting value\cite{Evans2011}. We believe that the Laplacian distribution of velocity components in our system arises due to the stochastic resetting seen in the particle's dynamics.

\subsection{Chirality}
We analyze the chirality in the dynamics, first by fitting a circle to the trajectory between time $t$ and $t+\Delta t$ ($\Delta t=0.16$ s) to obtain the center $(x_c,y_c)$ and radius $R$ of curvature. We find that the probability distribution $p(R)$ shows power-law behavior in contrast to most other circle swimming systems, which show a well-defined radius value determined by the anisotropy in the particle geometry\cite{PhysRevLett.110.198302}. For $\Gamma<\Gamma_c$, we find $p(R)\sim R^{-2}$, a behavior consistent with the $p(R)$ observed in our simulations of random-walk (see SI). On the other hand, for $\Gamma>\Gamma_c$, $p(R)$ shows, in general, three distinct regimes. For large $R$: $p(R)\sim R^{-2}$; for intermediate values of $R$: $p(R)\sim R^{-1}$ and for small values of $R$: $p(R)$ first flattens out and then begins to decrease as we decrease $R$. Further, the value $R_0$ where $p(R)$ transitions from a slope of $-1$ to  $-2$ increases with $\Gamma$, resulting in a shifting of probability to higher $R$ and a corresponding monotonic increase of the average radius $\langle R\rangle$ (see SI).

\begin{figure}
    \includegraphics[width=0.5\textwidth]
    {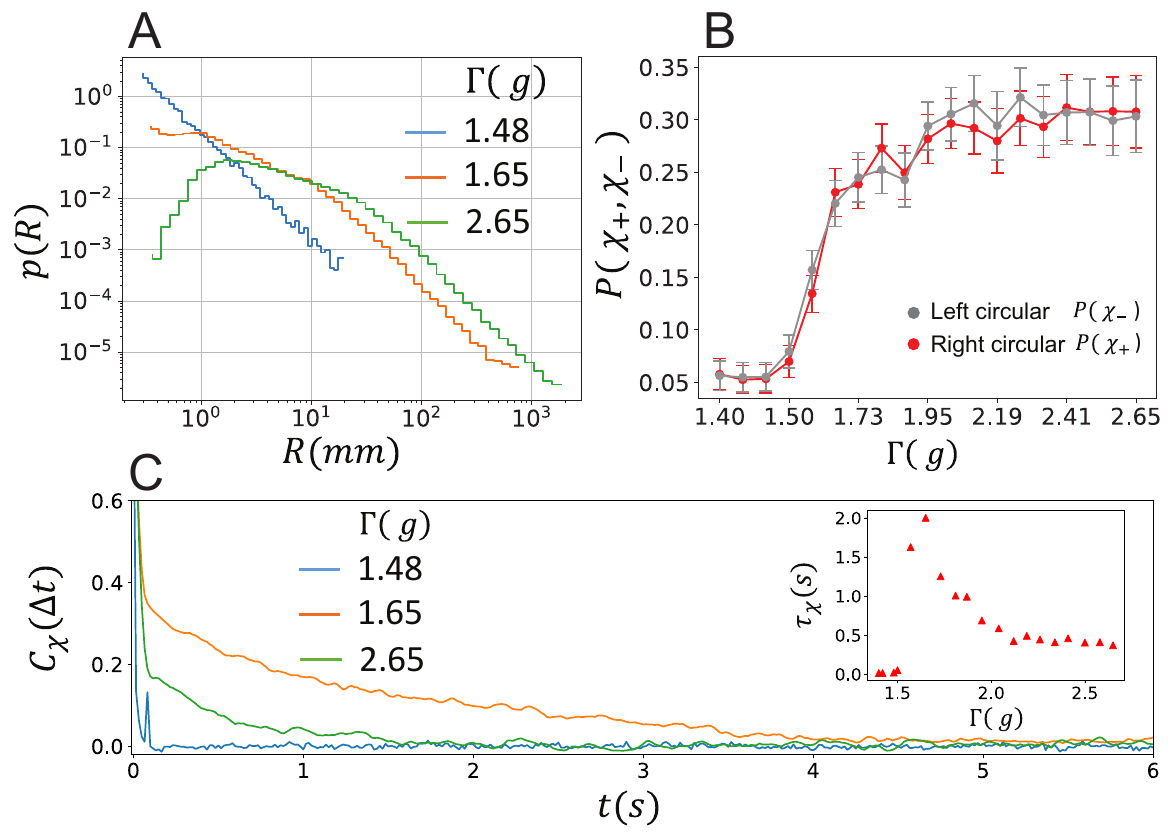}
    \caption{Curvature and chirality. (A) Probability distribution $p(R)$ of the radius of curvature of the trajectory for three different values of $\Gamma$. The distribution $p(R)$ shows a cross-over from $\sim 1/R$ to  $\sim 1/R^2$ for $\Gamma>\Gamma_c$. (B) Fraction of total time spent in right-circular trajectory $P(\chi_+)$ and left-circular trajectory $P(\chi_-)$ as a function of $\Gamma$.  (C) The auto-correlation function of $\chi$: $C_\chi (\Delta t)$. Inset shows the correlation time  $\tau_\chi (\Gamma)$.}

    \label{fig:chirality}
\end{figure}

The power-law distribution of $p(R)$ with cross-over in exponents is another indication that the system exists in a mixed state involving both random achiral as well as ordered chiral kinetics. We distinguish between different kinetic states by using the following order parameter to measure chirality: $\vec{\chi}=\hat{R}\times\hat{v}$, where $\hat{v}=\vec{v}/v$ and $\hat{R}=((x-x_c)\hat{x}+(y-y_c)\hat{y})/R$. 
The vector $\vec{\chi}=\chi\hat{z}$ points along the vertical direction, with $\chi$ lying between $1$ and $-1$. A value of $\chi$ close to $1$ implies a right-circular (anti-clockwise) trajectory, a value close to $-1$ a left-circular (clockwise) trajectory, and $\lvert \chi\rvert <1$ a random achiral trajectory. In Fig. \ref{fig:chirality}B, we plot the probability of right-circular $P(\chi_+)$ and left-circular $P(\chi_-)$ trajectory, as a function of $\Gamma$. Both $P(\chi_+)$ and $P(\chi_-)$ have negligibly small values for $\Gamma<\Gamma_c$ and increase rapidly for $\Gamma>\Gamma_c$ approaching a value close to $1/3$ in the limit of large $\Gamma$. For large $\Gamma$, the system is equally likely to be in the right-circular, left-circular, or random achiral state. This further establishes the trichotomous nature of the system. 

The parity symmetry in our system gets broken spontaneously at small time scales, but is reestablished at longer times and where we find that $P(\chi_+)=P(\chi_-)$(Fig. \ref{fig:chirality}B). We capture the evolution of the particle's kinetics from being chiral at short time scales to becoming achiral at longer times by computing the auto-correlation of $\chi$: $C_\chi(\Delta t)=\langle \chi(0)\chi(\Delta t)\rangle$ (Fig. \ref{fig:chirality}C). The auto-correlation $C_\chi(\Delta t)$ starts with a value close to $1$ and decays to zero over a time scale dependent on $\Gamma$. We obtain a measure of the correlation time  $\tau_\chi$ of $\chi$ by noting down the value of $\Delta t$ at which $C_\chi(\Delta t)$ reduces to $0.1$. We find that $\tau_{\chi}$ is negligibly small for $\Gamma<\Gamma_c$, increases rapidly first for $\Gamma>\Gamma_c$, and then decays slowly to saturate at $\tau_\chi\approx 0.5$ s for large $\Gamma$ (Fig. \ref{fig:chirality}C inset).

\section{Conclusion}
In conclusion, we have presented a simple granular system that shows chiral activity with circular particle trajectory. The particle has the commonplace shape of a torus, entailing an axially symmetric geometry and, indeed, for small vibration amplitude the system shows isotropic random motion. However, beyond a threshold vibration amplitude, it undergoes a transition driven by spontaneous symmetry breaking at multiple levels leading to persistent circular motion. First, as the particle orients itself at a large angle with the horizontal, its projected shape on the horizontal plane becomes elliptical, breaking the isotropy in the horizontal plane. Then, the fore-aft symmetry of the elliptical projection is broken kinetically due to the rolling motion of the particle. Further, while rolling persistently in one direction, the particle is inclined at some angle with the vertical, which amounts to a breaking of chiral symmetry leading either to right-circular or left-circular trajectory. 

The statistical mechanics of the particle's motion is distinctly different from other systems where activity and chirality arise from explicit asymmetries in the particle design. We see three kinetic states--two active chiral ones, with right-circular and left-circular orbits, and a passive achiral one. We observe transitions between all pairs of states. As a result, ordered chiral motion persists over macroscopic but finite time-scales and decays into a symmetric achiral form when averaged over longer times. In the limit of large driving amplitudes, the three states occur with equal probability, putting an upper bound on the time spent in chiral orbits. However, the transition between the achiral and the chiral states do not happen symmetrically and the temporal evolution of the system shows stochastic resetting. It is one of the few experimental systems showing stochastic resetting in a natural way and can serve as a model system to study the non–equilibrium physics and stochastic thermodynamics of chiral activity arising out of spontaneous symmetry breaking, both at single particle level as well as in the collective behavior. Moreover, the simple and robust self-propulsion mechanism can find applications in designing locomotion strategies in modern robotics.

\section*{Acknowledgement}
D.K. acknowledges the New Faculty Seed Grant, and Equipment Matching Grant from the Indian Institute of Technology (IIT) Delhi, and SRG/2019/000949 from the Science and Engineering Research Board (SERB), India for financial support. S.S. thanks IIT Delhi for institute fellowship. D.K. acknowledges support from the Department of Physics, IIT Delhi, in setting up a new research lab. S.S. and D.K. thank the Central Research Facility and Nanoscale Research Facility, IIT Delhi, for access to their facilities.

\bibliography{references}

\end{document}